\documentclass[a4paper,11pt]{article}
\usepackage{amssymb,amsmath,bm}  
\usepackage{graphics,graphicx}   
\usepackage{array,booktabs}      
\usepackage{authblk}  

\usepackage{cite}
\usepackage[margin=2.5cm]{geometry}

\usepackage[linktocpage,
colorlinks = true,
linkcolor = blue,
urlcolor  = blue,
citecolor = red,
anchorcolor = blue]{hyperref}

\usepackage{epsfig}

\numberwithin{equation}{section}

\title{\textbf{DAMPE Excess from Leptophilic Vector Dark Matter: Model Independent Approach}}
\author[1]{Seyed Yaser Ayazi\thanks{syaser.ayazi@semnan.ac.ir}}
\author[2,3]{Ahmad Mohamadnejad\thanks{mohamadnejad.a@lu.ac.ir}}
\affil[1]{Physics Department, Semnan University, P.O. Box. 35131-19111, Semnan, Iran}
\affil[2]{Department of Physics, Lorestan University, khorramabad, Iran}
\affil[3]{School of Particles and Accelerators,
Institute for Research in Fundamental Sciences (IPM), Tehran, Iran} 
\date{\today}

\begin{document}

\baselineskip 0.6 cm

\maketitle

\begin{abstract}
We study all extensions of the Standard Model (SM) with a vector dark matter (VDM) candidate which can explain the peak structure observed by recent DAMPE experiment in electron-positron cosmic ray spectrum.
In this regard,  we consider all leptophilic renormalizable VDM-SM interactions through scalar, spinor, and vector mediators.
We show that only two out of six possible models could produce DAMPE signal by annihilation of VDM with the mass about 1.5 TeV in a nearby subhalo whilst simultaneously satisfying constraints from DM relic density, direct/indirect detection, and the collider bounds. These two models are the ones with scalar/pseudoscalar mediator $ \phi $ with $ M_{\phi} \in [1500,3000] $ GeV.
\end{abstract}



\section{Introduction} \label{sec1}
In the recent years, several of cosmic ray (CR) detectors in the space have presented a wide range of the new data ($1~\rm GeV$-$10~\rm TeV$) in the electron-positron spectra; the Fermi Large Area Telescope (Fermi-LAT) has reported a measurement of the CR electron-positron spectrum from $7~\rm GeV$ to $2~\rm TeV$\cite{Fermi-Lat}. The PAMELA satellite 
 experiment\cite{PAMELA} observed an abundance of the positron in the CR energy range of $15-100~\rm GeV$, also a positron fraction in primary CRs of $0.5-500~\rm GeV$ \cite{Accardo:2014lma} and the measurement of electron plus positron flux in the primary CRs from $0.5~\rm GeV$ to $1~\rm TeV$  have been reported by the Alpha Magnetic Spectrometer (AMS-02)\cite{AMS}. Also the results of a CR electron-positron spectrum, between $10~\rm GeV$ and $\rm 3 TeV$, have been presented based upon observations with the CALorimetric Electron Telescope instrument(CALET)\cite{CALET}. The recent report of the DAMPE collaboration \cite{TheDAMPE:2017dtc} released measurements of the electron-positron spectrum in the energy range $25~\rm  GeV$ to $4.6 ~\rm  TeV$ with high energy resolution and low particle background. Although DAMPE data confirm the measurements of the AMS-02 \cite{AMS} and Fermi-LAT \cite{Fermi-Lat} on the low energy side (below $\rm 1 ~TeV$), the spectrum seems to have a tentative sharp peak above the background at around $1.4 ~\rm  TeV$ \cite{TheDAMPE:2017dtc}. It is notable that in the energy region below $300~ \rm GeV$, CALET's spectrum is consistent with the AMS-02, Fermi-LAT, and DAMPE, while in the energy ranges $300$ to $1~ \rm TeV$ the CALET’s results exhibit a lower flux than the spectra from the latter two experiments\cite{Adriani:2018ktz}. While DAMPE excess could be a statistical fluctuation \cite{Fowlie:2017fya}, the extensive discussion on the possible theoretical and experimental explanation of the DAMPE excess” with  both astrophysical origin \cite{pulsar} or DM origin  \cite{DM}
have been proposed. The monoenergetic electron in the energy $1.4 \rm ~TeV$ implies local sources of electrons-positrons, because $\rm TeV$ electrons can only travel by a small distance $(\rm kpc)$ in the Milky Way due to strong radiative cooling process of high energy
CR electrons. Therefore, if this excess emanates from DM, the source of such high energy and monoenergetic electrons is located at the vicinity of the solar system\cite{Yuan:2017ysv}. One possible way to describe electron-positron excess is that the DM particles annihilate into leptons and the mass of DM particles are about $1.5 \rm ~TeV$ if the nearby DM subhalo of $\sim 10^7\,M_\odot$ locates $0.1-0.3 \rm~kpc$ away from solar system and the
DM annihilation cross section is $ \left\langle \sigma v \right\rangle \sim 3 \times 10^{-26} \, cm^{3}/s $\cite{Yuan:2017ysv}.

Here, we explain the DAMPE electron excess by attributing to the VDM annihilation in the near of the solar system. 
We have extended the study performed in \cite{Athron:2017drj} to cover VDM candidate. In this reference only the scalar and fermionic DM candidates was considered.
The VDM and some of its theoretical and phenomenological impacts has been extensively  studied in literatures\cite{VDM}. In this regard, we classify all renormalizable  VDM models with leptophilic interactions in which a massive particle  with spin 0, 1/2, or 1 plays the role of mediator between dark side and the SM leptons.

The rest of this paper is organized as follows: In the next section, we extend SM with the set of simplified leptophilic vector dark mater models that couple with  scalar, spinor or vector mediators. In
Sec.~\ref{sec3}, we introduce the conditions for explaining the DAMPE electron excess in the models. In this section, we also discuss phenomenological constraints such as anomalous magnetic moments of leptons, direct detection, indirect detection  and collider constraints on the parameter space of the models. The combined analysis for DAMPE excess and phenomenological constraints in parameters space are given in Sec.~\ref{sec4}. Finally, we make a conclusion in Sec.~\ref{sec5}.

\section{Vector dark matter models} \label{sec2}
In this paper, we consider a model-independent
approach in which we study all renormalizable interactions via a massive spin 0, 1/2, or 1 mediator between
VDM particles and the SM leptons.
In our study a single species of VDM is responsible for both DAMPE excess and the DM relic density. 
We study the following six possible interactions between VDM and SM leptons which satisfy Hermiticity, Lorentz invariance, and renormalizablity,
\begin{align}
{\text{model 1:}} \quad {\cal{L}}_{1} &\supset \mu \phi X_{\mu} X^{\mu} + \sum_{\ell=e,\mu,\tau} \lambda_{s} \phi \overline{\ell} \ell , \\
{\text{model 2:}} \quad {\cal{L}}_{2} &\supset \mu \phi X_{\mu} X^{\mu} + \sum_{\ell=e,\mu,\tau} \lambda_{p} \phi \overline{\ell} i  \gamma^{5} \ell , \\
{\text{model 3:}} \quad {\cal{L}}_{3} &\supset g_{v} V_{\mu} (X^{\nu} \partial\nu X^{\mu} + {\text{h.c.}}) + \sum_{\ell=e,\mu,\tau} g_{s}  V_{\mu} \overline{\ell} \gamma^{\mu} \ell , \label{lagrangian3}\\
{\text{model 4:}} \quad {\cal{L}}_{4} &\supset g_{v} V_{\mu} (X^{\nu} \partial\nu X^{\mu} + {\text{h.c.}}) + \sum_{\ell=e,\mu,\tau} g_{p} V_{\mu} \overline{\ell}  \gamma^{\mu} \gamma^{5} \ell , \label{lagrangian4}\\
{\text{model 5:}} \quad {\cal{L}}_{5} &\supset \sum_{\ell=e,\mu,\tau} y_{s} X_{\mu} \overline{\psi} \gamma^{\mu} \ell + {\text{h.c.}} , \quad M_{\psi} > M_{X} , \\
{\text{model 6:}} \quad {\cal{L}}_{6} &\supset \sum_{\ell=e,\mu,\tau} y_{p} X_{\mu} \overline{\psi} \gamma^{\mu} \gamma^{5} \ell + {\text{h.c.}} , \quad M_{\psi} > M_{X} ,
\end{align}\label{lagrangian}
where $ X_{\mu} $ is the VDM candidate, and, $ \phi $, $ V_{\mu} $, and $ \psi $ are scalar, vector, and (Dirac) spinor mediators, respectively. In our analysis, we have assumed universal couplings between the mediators and the SM leptons. To keep it simple, we have avoided mixing between generations. 
We also define the dimensionless
coupling $ g_{\phi} = \frac{\mu}{M_{X}} $, so that all parameters be dimensionless.

In the case of spin 0 mediators (model 1 and 2), couplings between the scalar field and only left-handed neutrinos are zero. For the right-handed neutrinos, considering see-saw mechanism, Yukawa couplings
between light mass eigenstates and the scalar mediator would be suppressed by lightness of neutrino masses or equivalently the seesaw scale. Since $ X_{\mu} $ is neutral with no electric charge, for spin 1/2 mediators (model 5 and 6), the spinor $ \psi $ has positive electric charge (and couples to photon), that is, equal but opposite to the charged leptons. Therefore, $ X_{\mu} $ and neutrinos can not couple together via charged spin 1/2 mediators. These models are also considered in \cite{Feng:2019rgm} to explain the positron and electron flux from
AMS-02 and DAMPE data. Unlike our work, they have considered both DM annihilation and decay into light leptons at the same time.
Since in any SU(2) invariant theory, the coupling between neutrinos and the vector mediator is generally non-zero in the case of spin 1 mediators (model 3 and 4), the vector mediator, $ V_{\mu} $, can couple to the both charged leptons and left-handed neutrinos
by a vector or axial-vector interactions. However, considering \eqref{lagrangian3} and \eqref{lagrangian4}, models 3 and 4 will lead to a $p$-wave suppressed annihilation cross section and they can not explain DAMPE excess, therefore, there is no need to consider Yukawa interactions between the vector mediators and neutrinos in these models.

In each one of models 1 to 6, we only consider a single mediator and a single species of VDM in which the lepton-mediator interactions is either completely scalar (vector) or completely pseudoscalar (axial-vector), but not a mixture. As we mentioned before, we assume universal couplings only between charged leptons and the mediators.
Furthermore, no tree-level mixing between the SM Z-boson and VDM has been assumed. We have also ignored tree-level mixing between $ \phi $ and the SM Higgs boson. Also note that for models 5 and 6, in order to avoid DM decay, we should have $ M_{\psi} > M_{X} $. Regarding this constraint $ X_{\mu} $ will be stable and can serve as DM.
\section{Phenomenological constraints} \label{sec3}
In this section, we study various constraints on parameters space of the model so that it predicts the correct relic abundance of DM and the DAMPE excess. These constraints coming from experimental observables at LEP and LHC and DM direct/indirect detection. These are furnished in the following.

\subsection{DAMPE electron-positron excess} 
DAMPE measurements of the cosmic electron-positron flux exhibit a sharp resonance near 1.4 TeV which hints DM annihilation (or decays) in a DM subhalo located close to the solar system with an enhanced DM density.
Because this sharp resonance in the DAMPE data
occurs around 1.4 TeV, we take the VDM mass to be 1.5 TeV. In order to produce DAMPE peak, one requires DM subhalo with a density about 17-35 times greater than the local density of DM at a distance of $ \sim $ 0.1 kpc \cite{Yuan:2017ysv}. Moreover, DM annihilation cross section should not be suppressed by velocity ($ \sigma v \sim v^{2} $). In models 1 to 6, we take $ \left\langle \sigma v \right\rangle \simeq 
[2.2-3.8] \times 10^{-26} \, cm^{3}/s $ as a constraint required to explain DAMPE excess.

\subsection{Relic density} 
\begin{figure}[!htb]
\begin{center}
\centerline{\hspace{0cm}\epsfig{figure=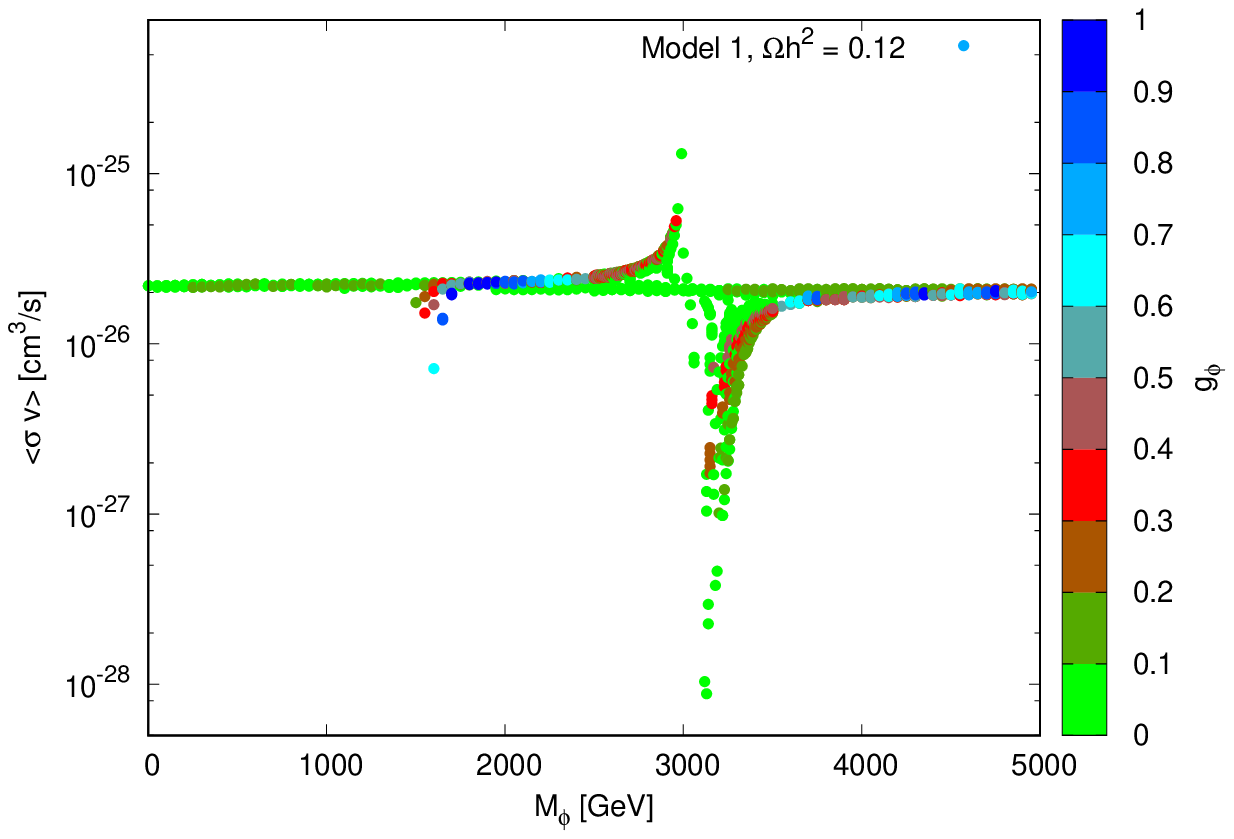,width=7.5cm}\hspace{0cm}\epsfig{figure=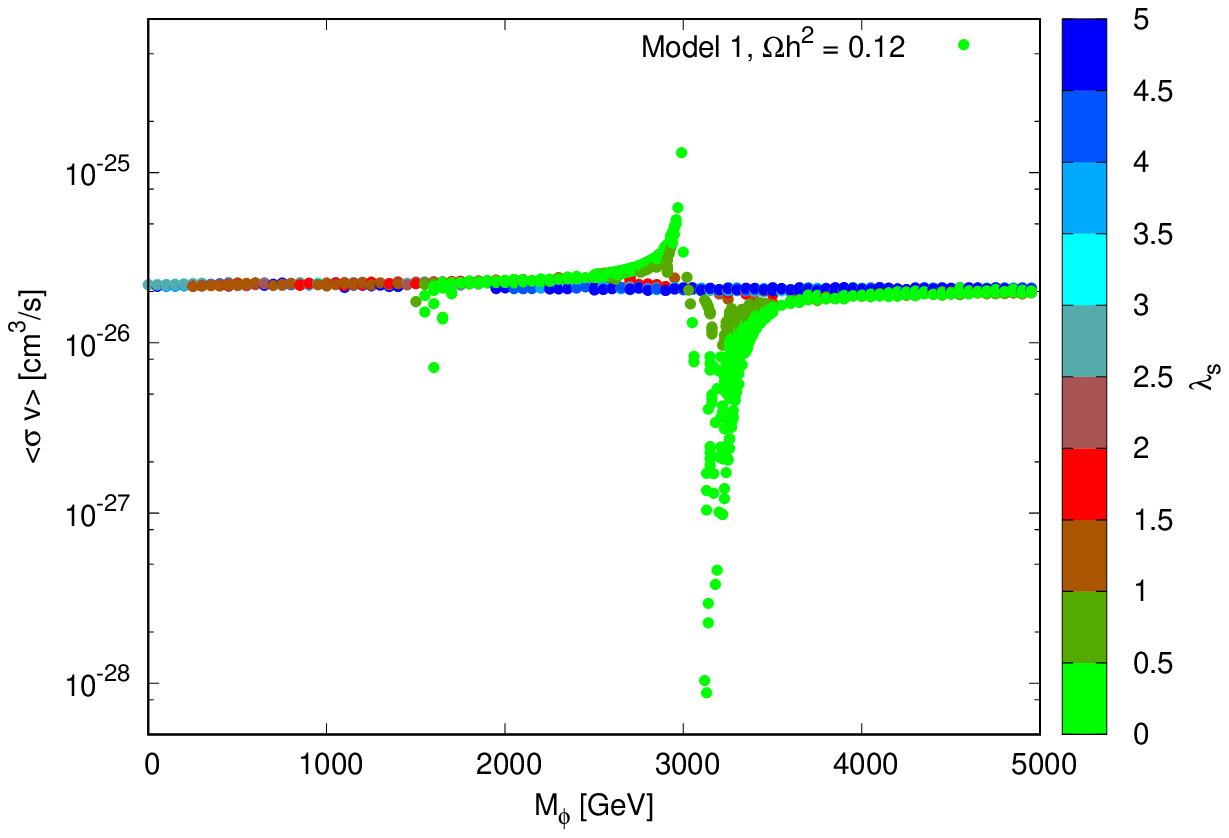,width=7.5cm}}
\centerline{\hspace{0cm}\epsfig{figure=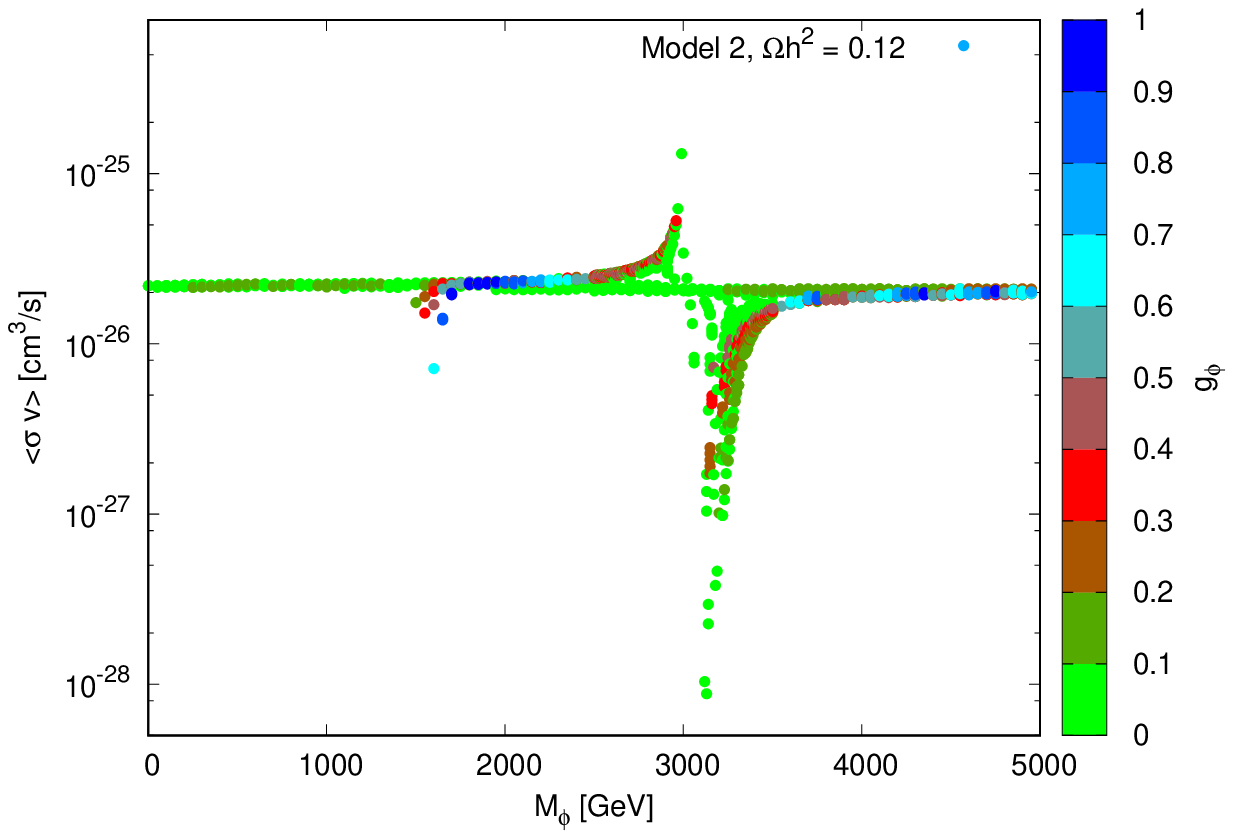,width=7.5cm}\hspace{0cm}\epsfig{figure=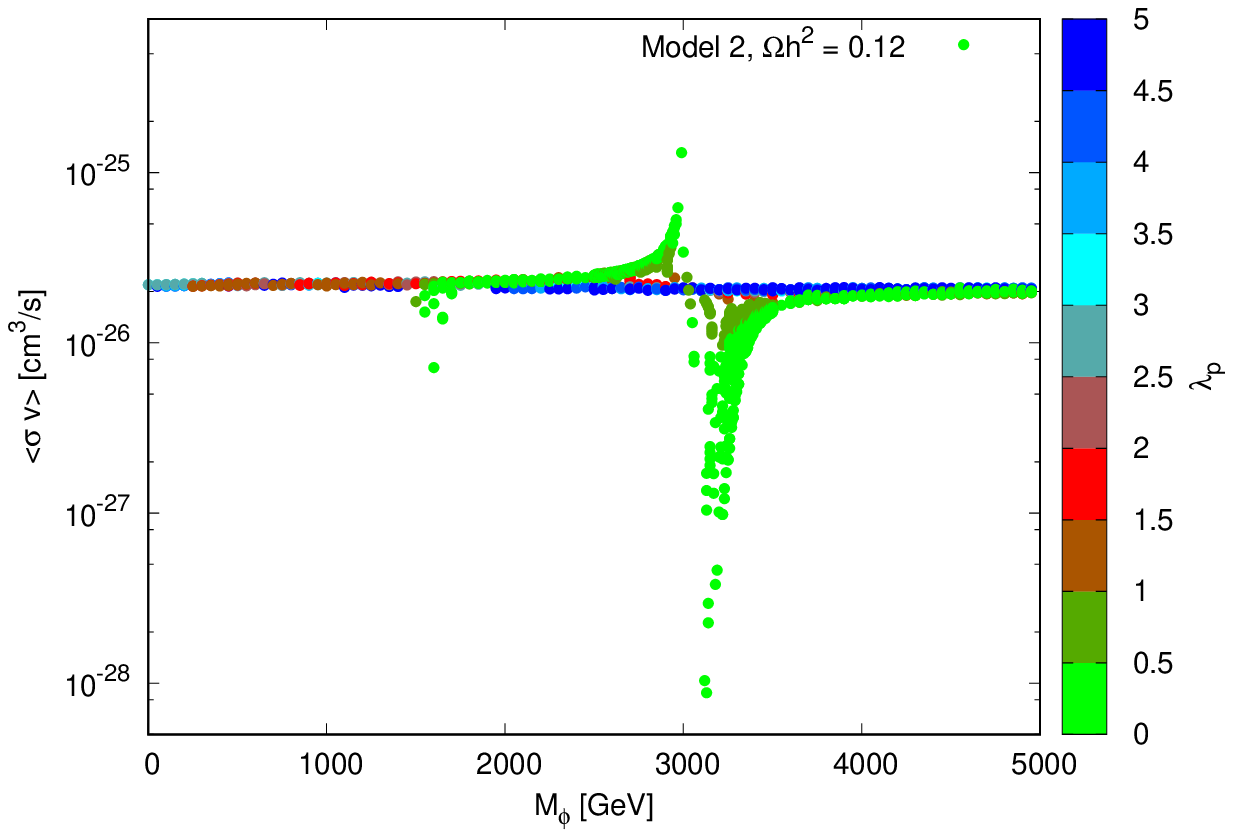,width=7.5cm}}
\centerline{\hspace{0cm}\epsfig{figure=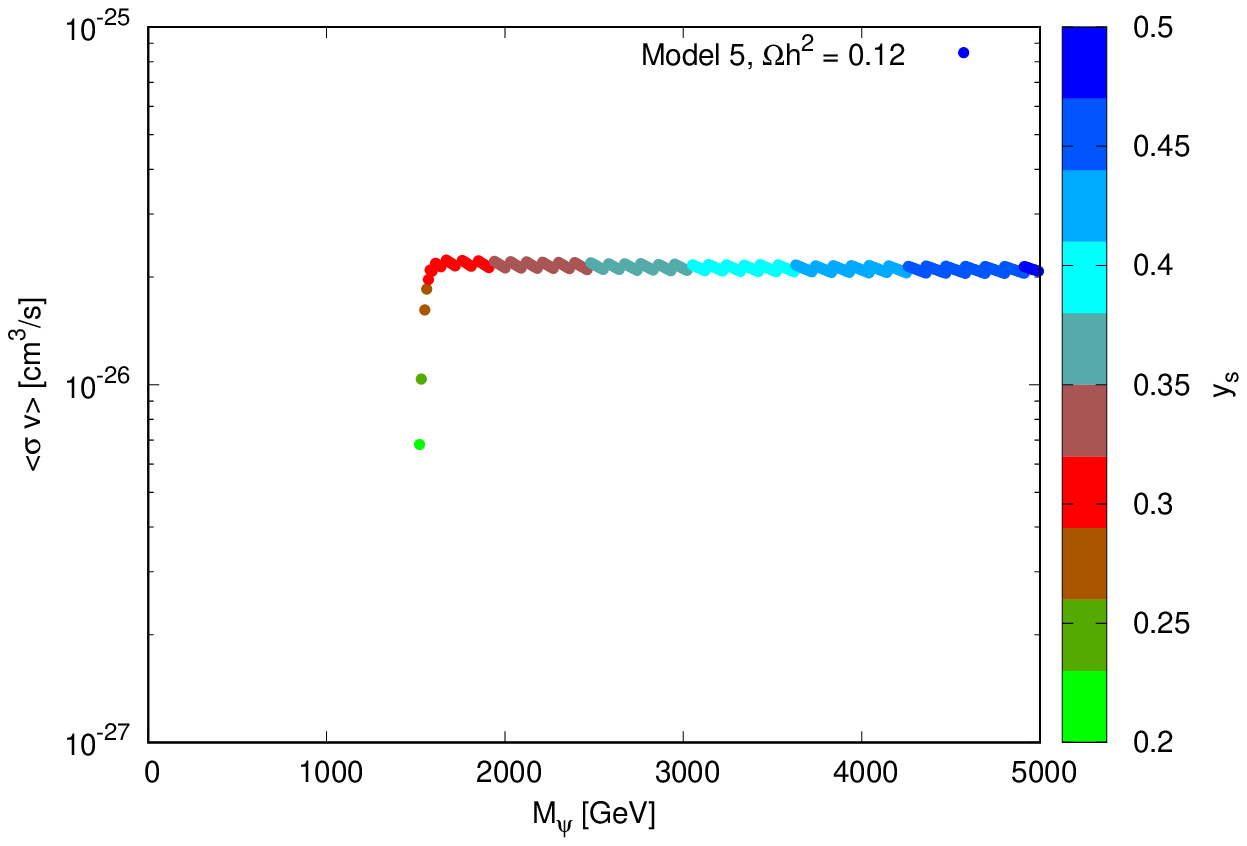,width=7.5cm}\hspace{0cm}\epsfig{figure=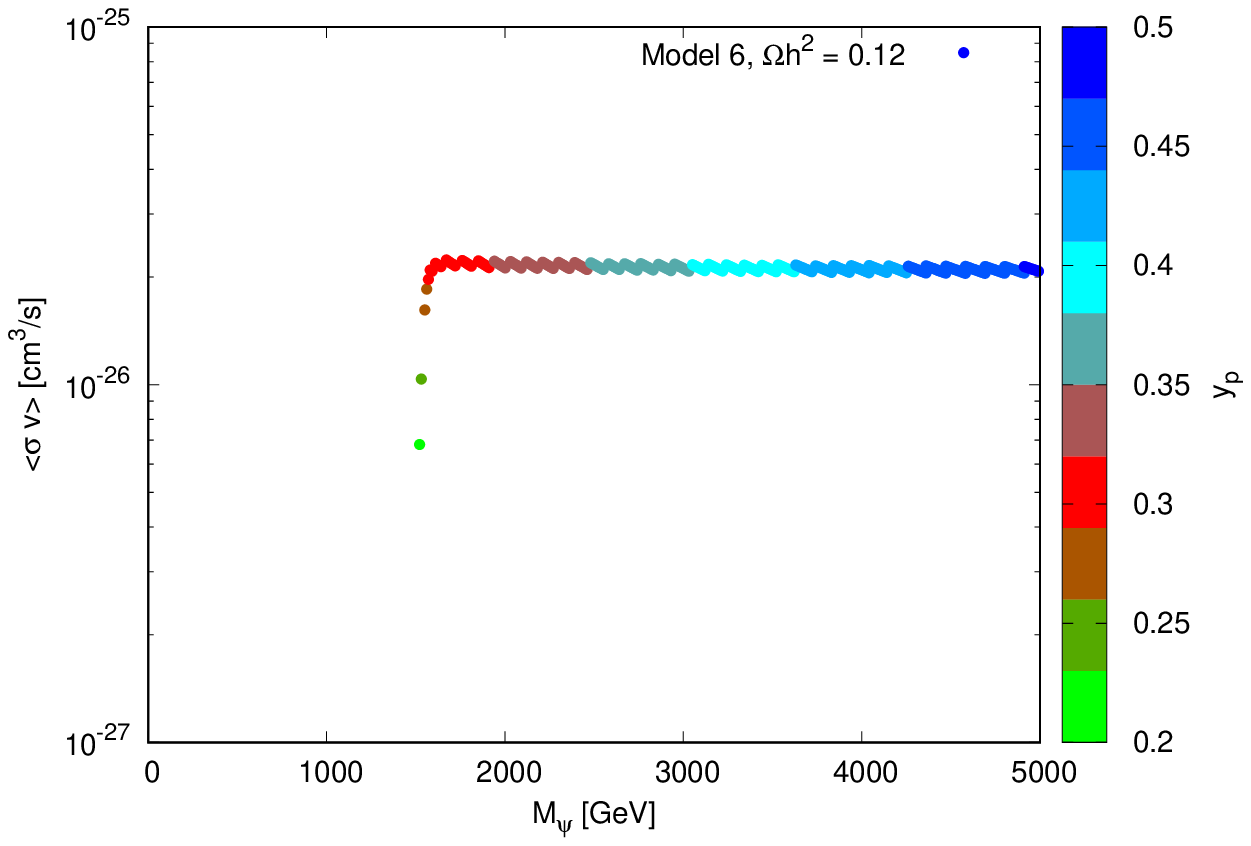,width=7.5cm}}
\centerline{\vspace{-0.7cm}}
\caption{Total cross section of DM annihilation at the present Universe versus the mediator mass for the parameters space of the models 1, 2 , 5 and 6 which are consistent with DM relic density. In this figure the DM mass is 1.5 TeV.} \label{ann}
\end{center}
\end{figure}

\begin{figure}[!htb]
\begin{center}
\centerline{\hspace{0cm}\epsfig{figure=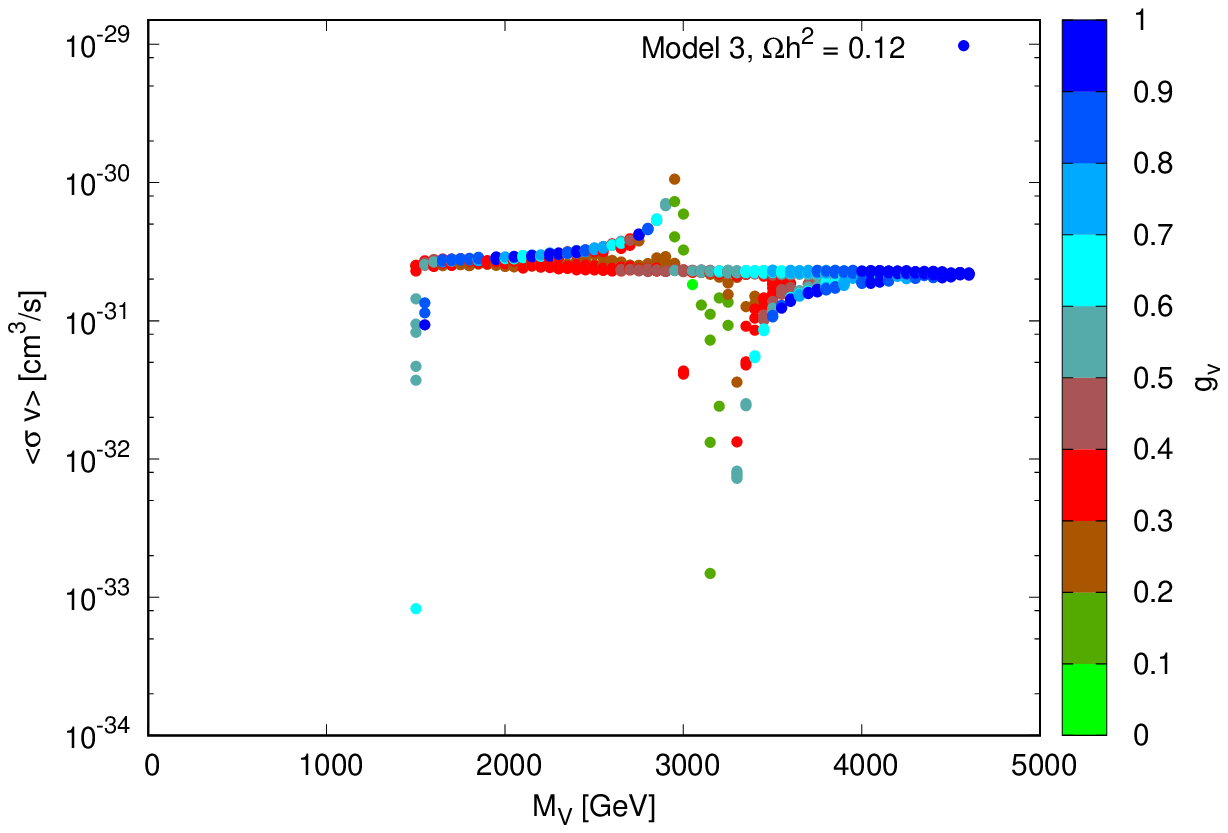,width=7.5cm}\hspace{0cm}\epsfig{figure=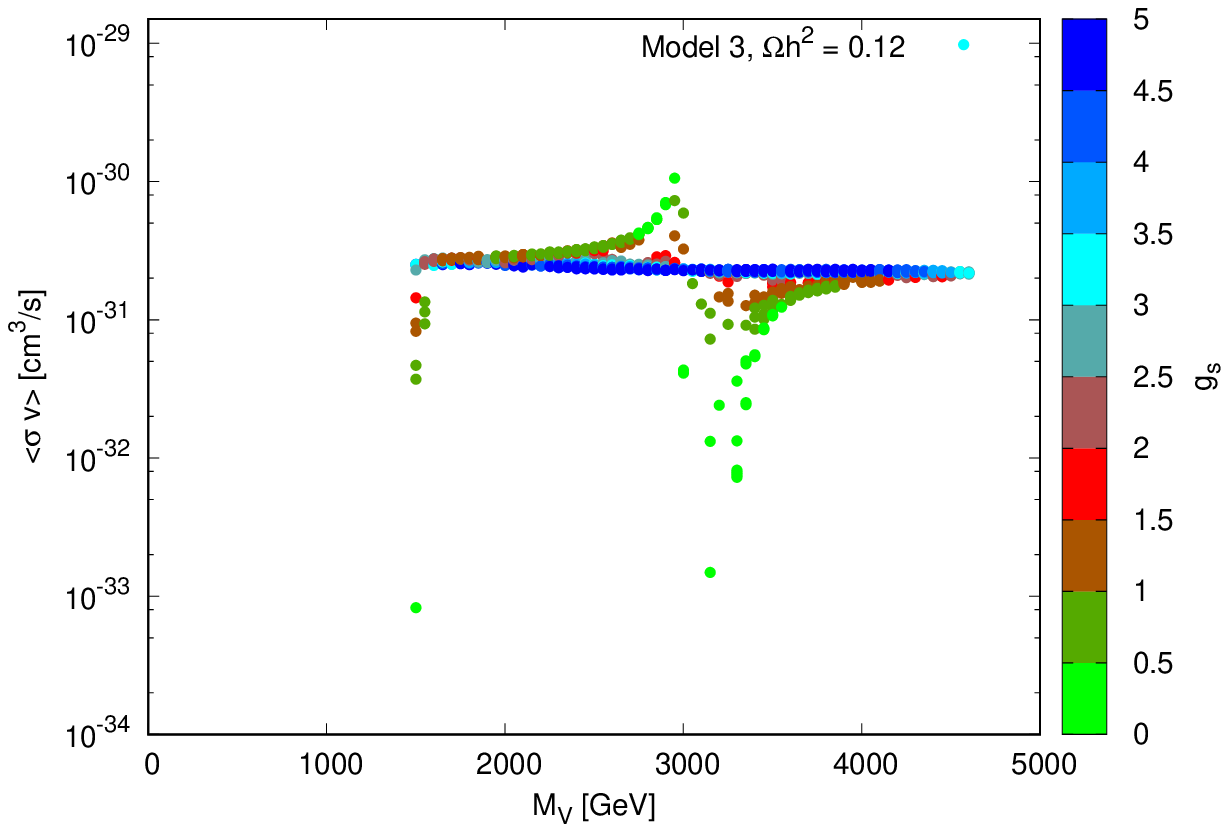,width=7.5cm}}
\centerline{\hspace{0cm}\epsfig{figure=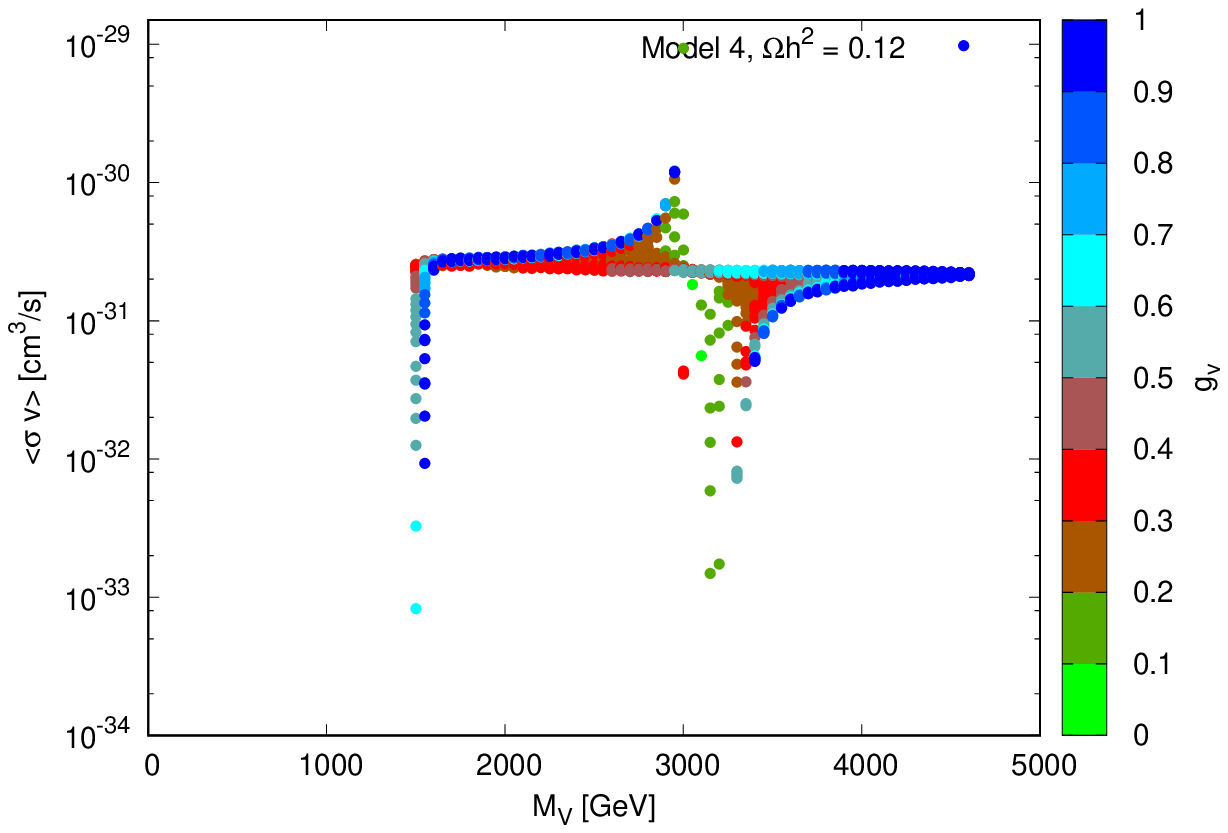,width=7.5cm}\hspace{0cm}\epsfig{figure=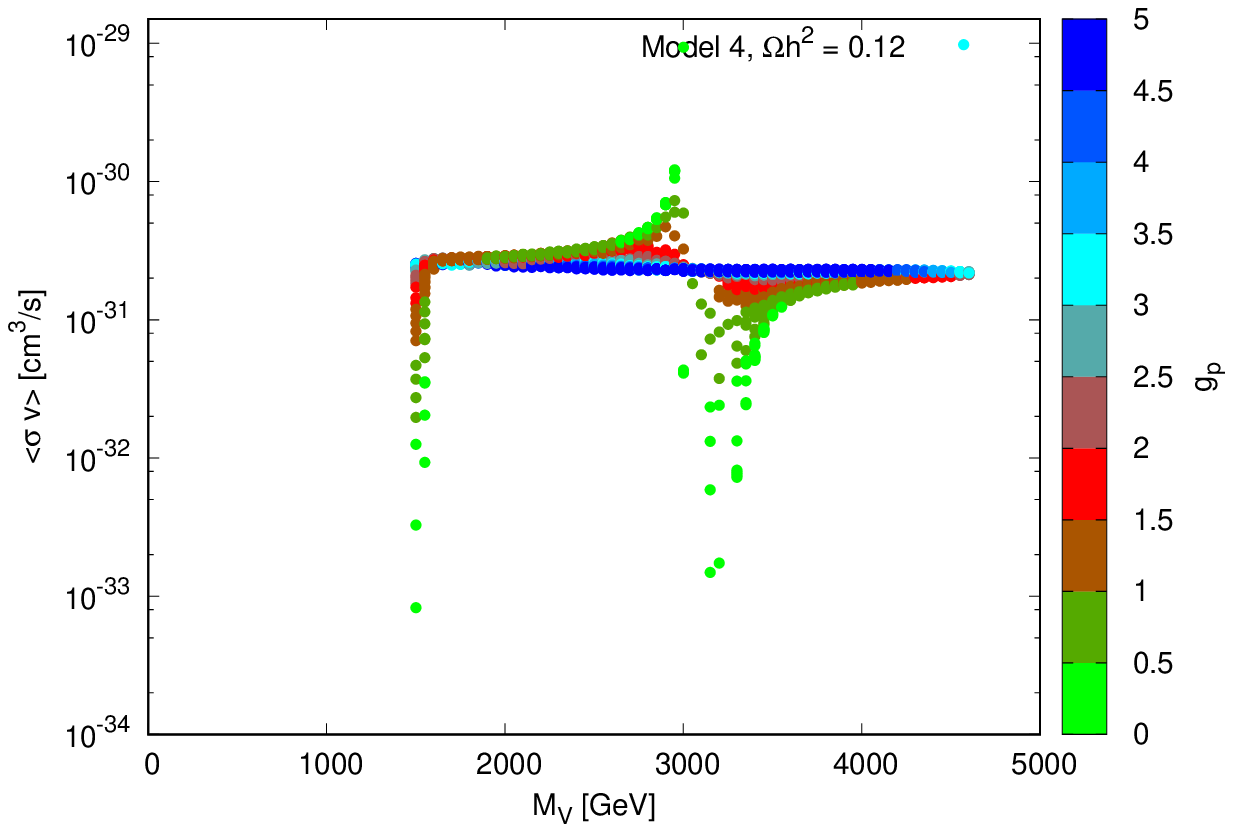,width=7.5cm}}
\centerline{\vspace{-0.7cm}}
\caption{Same as Fig.~\ref{ann} for models 3 and 4} \label{ann2}
\end{center}
\end{figure}

Here,
we compute the relic density for VDM candidate. According to Planck collaboration DM relic density is $ \Omega h^{2} = 0.120 \pm 0.001 $ \cite{Aghanim:2018eyx} which translates
into a strict relation between the couplings and mediator masses. In Figs.~\ref{ann} and \ref{ann2}, we depict DM annihilation cross section against the mediator mass for the parameters satisfying DM relic density. We have obtained DM relic density and annihilation cross section using {\tt{micrOMEGAs}} public code \cite{Barducci:2016pcb}. For models 3 and 4, DM annihilation cross section will be suppressed by velocity and we obtain $ \sigma v \simeq 10^{-31} \, cm^{3}/s $, which excludes the whole parameter space. In these models, the s-wave DM annihilation to leptons is absent. Hence, as the p-wave term is suppressed by a factor of the DM velocity squared, the annihilation cross section is not large enough to produce DAMPE signal. Note that only the parameter space for which the contribution of DM annihilation to leptons is more than 30 percent is depicted. However, for models 1, 2, 5, and 6 only a small region of the parameter space will be excluded.

In Fig.~\ref{ann}, for models 1 and 2 there are two noticeable dips at $ M_{\phi} \sim M_{X} $ (at which annihilation proceeds through a t-channel resonance into $\phi\phi$) and at
$ M_{\phi} \sim 2 M_{X} $ (at which annihilation proceeds through an s-channel resonance into $l\bar l$.). In these cases
reduced couplings are required to get the relic density constrained by Planck data. As it is seen in Fig.~\ref{ann} and \ref{ann2}, for s-channel resonance, there is an increase around $M_{\phi}\sim 3~ \rm TeV$  and after that there is a decrease in cross section of DM in present Universe. In these models, for DM annihilation into leptons through s-channel, we have:
\begin{equation}
\sigma v \approx \frac{a}{(M_{\phi}^{2}-4 M_{X}^{2})^2} + 
\frac{b}{(M_{\phi}^{2}-4 M_{X}^{2})^3} v^{2} ,\label{cross-section}
\end{equation}
here $ v $ is the speed of the DM and $ a $ and $ b $ depend on the parameters of the model (for explicit form of $ a $ and $ b $ see Eq.~(B33) in \cite{Berlin:2014tja}). 
Regarding $ M_{X} = 1.5 $ TeV, $ a $ and $ b $ take positive values.
In non-resonance regions, the first term in \eqref{cross-section} dominates and
\begin{equation}
\left\langle \sigma v \right\rangle_{freeze-out} \approx \left\langle \sigma v \right\rangle_{today} \approx \frac{a}{(M_{\phi}^{2}-4 M_{X}^{2})^2} .\label{today}
\end{equation}
However, one cannot ignore the second term in 
 \eqref{cross-section} near the resonance. Therefor, in this region we have Eq. \eqref{cross-section} at freeze-out and Eq. \eqref{today} at the present Universe (with $ v=0 $). Consequently, if $ M_{\phi} < 2 M_{X} $, then $ \left\langle \sigma v \right\rangle_{today} > \left\langle \sigma v \right\rangle_{freeze-out} $, and if $ M_{\phi} > 2 M_{X}, $ then $ \left\langle \sigma v \right\rangle_{today} < \left\langle \sigma v \right\rangle_{freeze-out} $.
 This feature can be seen clearly in Fig.~\ref{ann}. (In \eqref{cross-section} widths of the scalar mediator is not included. Considering this the cross section will not be infinite at the resonance point. However, this does not affect our argument here.) Note that, in ref \cite{Athron:2017drj} there is only one dip where annihilation proceeds through an s-channel resonance. 
 Indeed, it seems in ref \cite{Athron:2017drj} only DM annihilation into leptons through s-channel is considered and they did not examine the other channel, i.e., t-channel annihilation into scalars. Both of this annihilation channels contribute in DM annihilation and are relevant in calculation of DM relic density. In ref \cite{Athron:2017drj} the authors considered the multiplication of the couplings between mediator-DM and mediator-leptons as the only relevant parameter, which makes sense in the case of DM annihilation into leptons through s-channel.
However, both DM annihilation into leptons and scalar mediators should be considered simultaneously. In the t-channel annihilation only the coupling between mediator-DM would be relevant.
Since we have included the t-channel annihilation in calculating DM relic density, these two couplings will be disentangled, and the annihilation cross-sections depend on both of them separately.
Furthermore, in this reference, mediator masses around s-channel resonance are excluded, while in our calculations this region is consistent with $ \left\langle \sigma v \right\rangle \simeq 
[2.2-3.8] \times 10^{-26} \, cm^{3}/s $ and could explain the DAMPE excess.

\subsection{Anomalous magnetic moments of leptons}  
In this section, we investigate constraints on the parameter space of the models which are imposed by the measurement of anomalous magnetic moments (AMM)s of leptons. Since in our models, VDM particles interact with the SM leptons via a massive mediator, a significant effect on the anomalous magnetic moments of the leptons is expected.
As it was mentioned in Sec.~\ref{sec2}, we suppose a lepton universal coupling for all interactions. Therefore, we consider only the magnetic moment of the muon and ignore weaker constraints on the (AMM)s of tau and electron.

The prediction for the value of the muon AMM in SM includes the contributions from virtual QED, electroweak, and hadronic processes. While the QED and electroweak processes account for most of the anomaly, the hadronic uncertainty cannot be calculated accurately from theory alone. It is estimated from experimental measurements of the ratio of hadronic to muonic cross sections in electron-positron collisions\cite{Xiao:2017dqv}. In \cite{Giusti:2017jof}, it is shown that the measurement can be interpreted as an inconsistency with the SM and suggesting physics beyond the SM may be having an effect. 

\begin{figure}
\begin{center}
\centerline{\hspace{0cm}\epsfig{figure=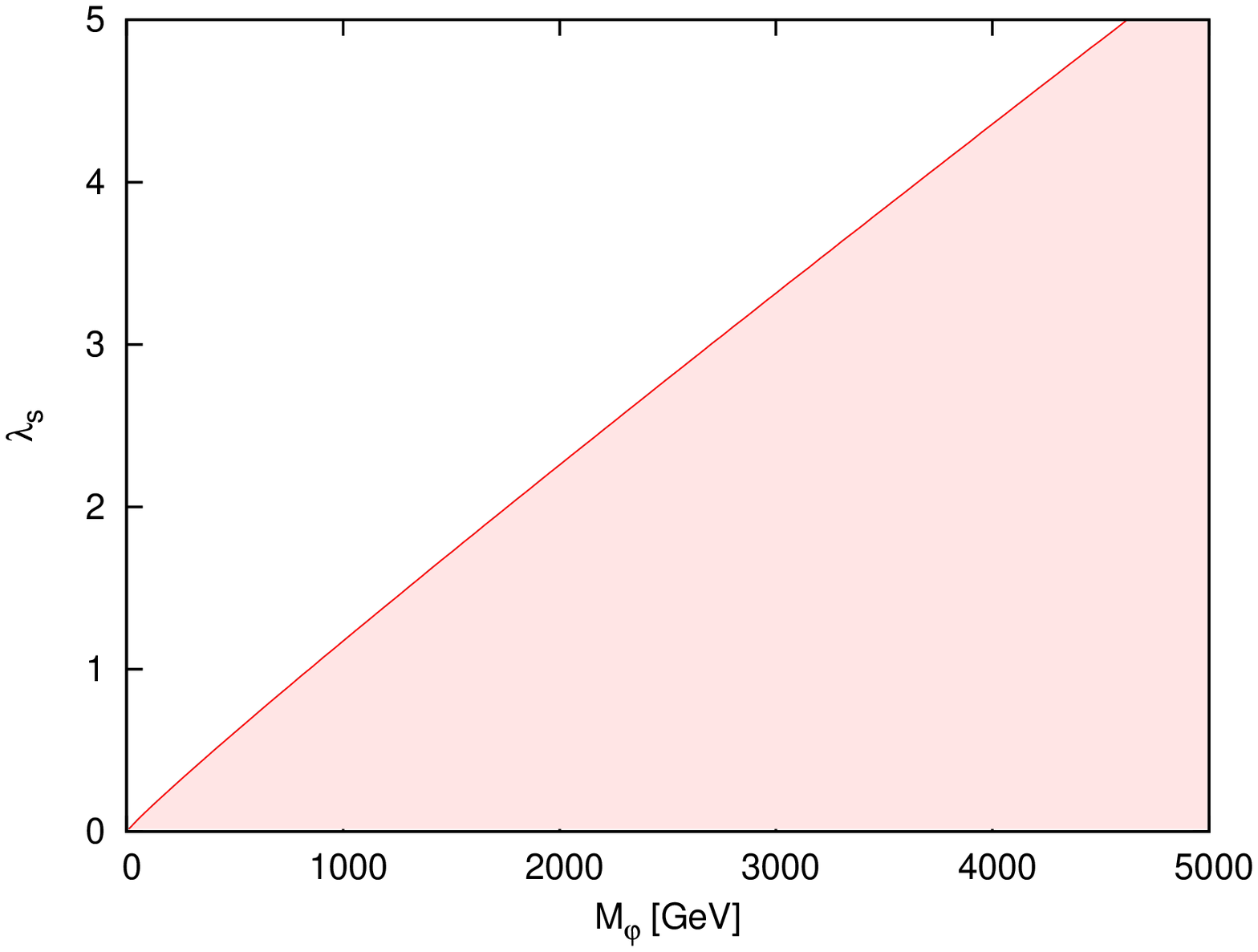,width=7cm}\hspace{0cm}\epsfig{figure=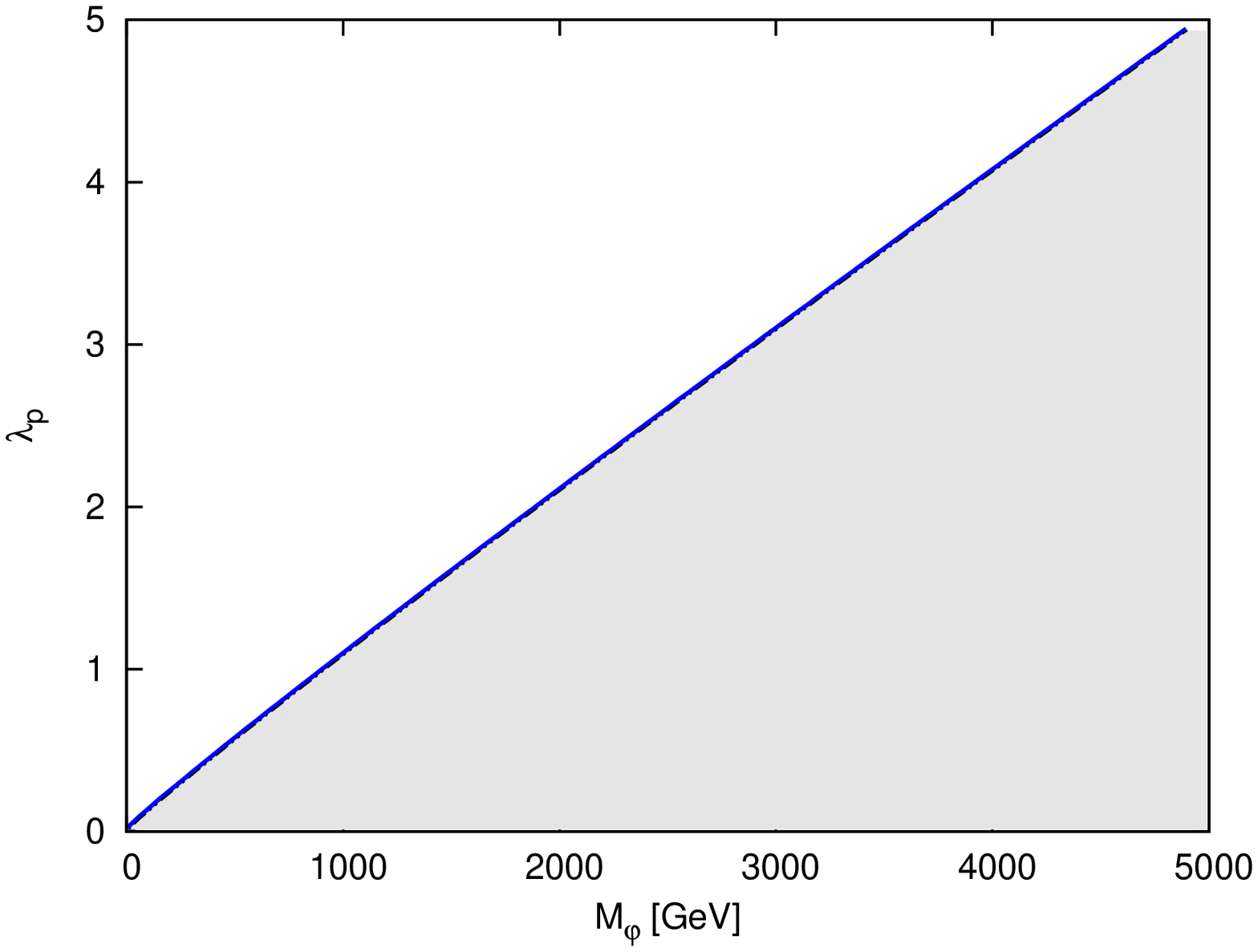,width=7cm}}
\centerline{\vspace{0.5cm}\hspace{0.5cm}(a)\hspace{6cm}(b)}
\centerline{\hspace{0cm}\epsfig{figure=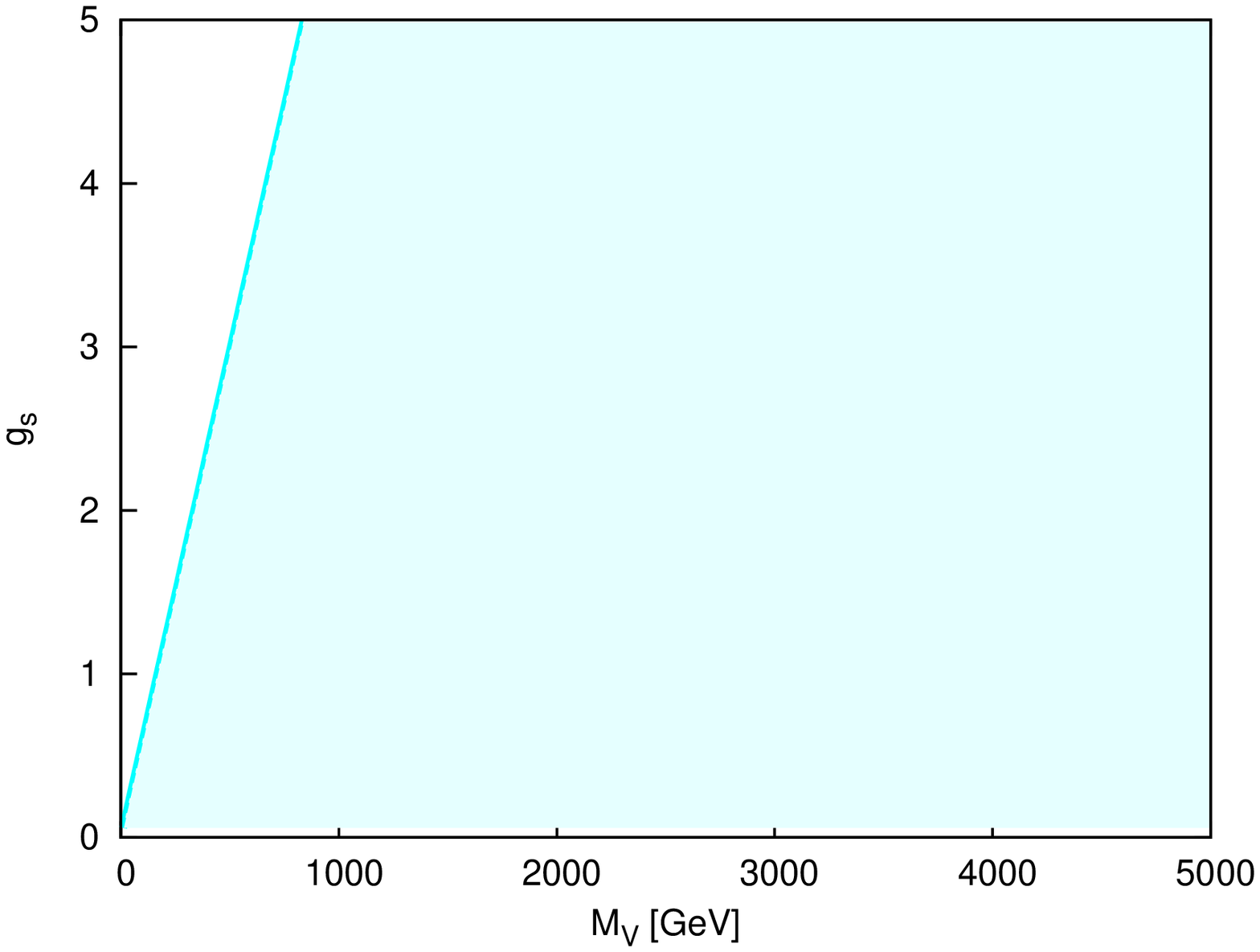,width=7cm}\hspace{0cm}\epsfig{figure=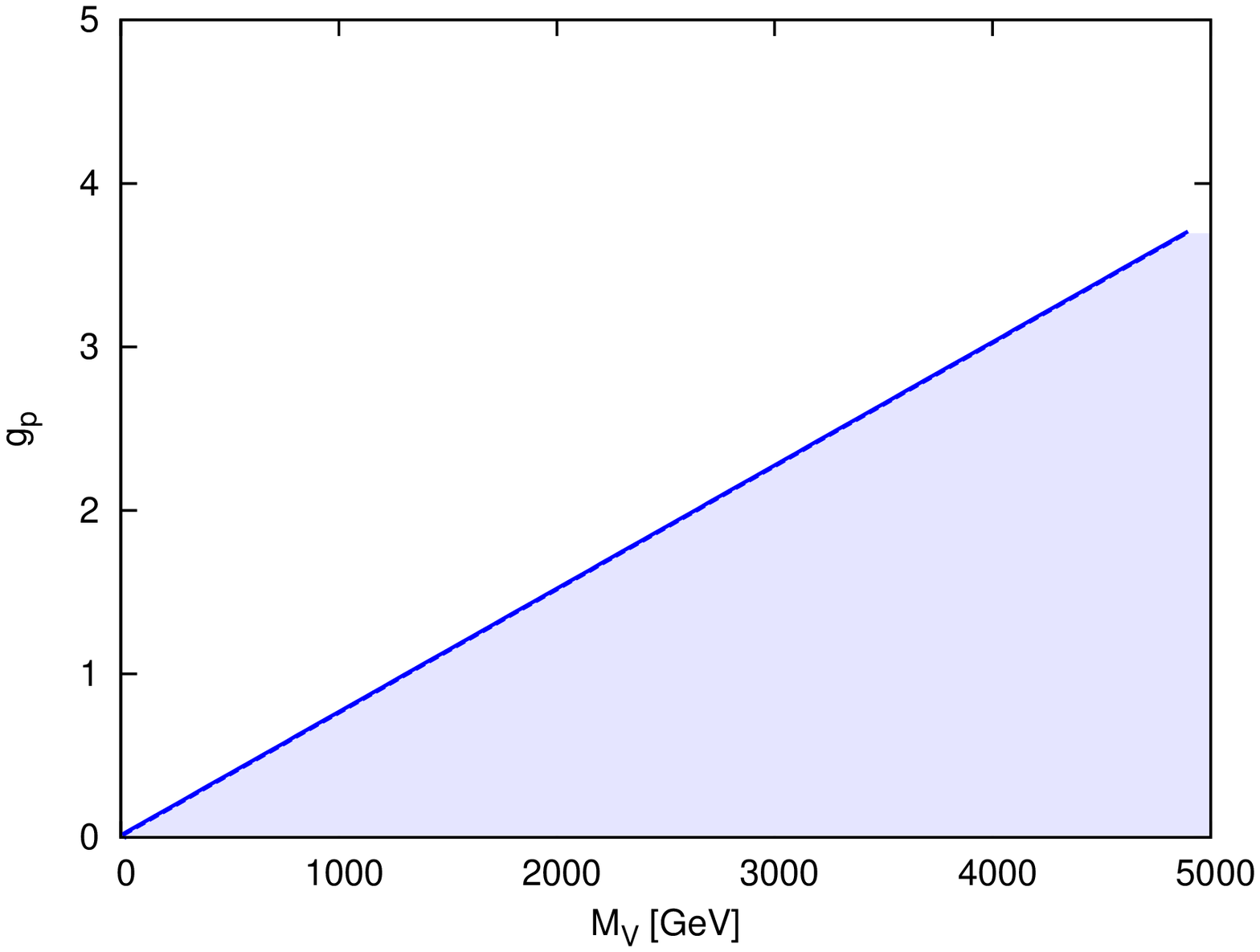,width=7cm}}
\centerline{\vspace{0.5cm}\hspace{0.5cm}(c)\hspace{6cm}(d)}
\centerline{\hspace{0cm}\epsfig{figure=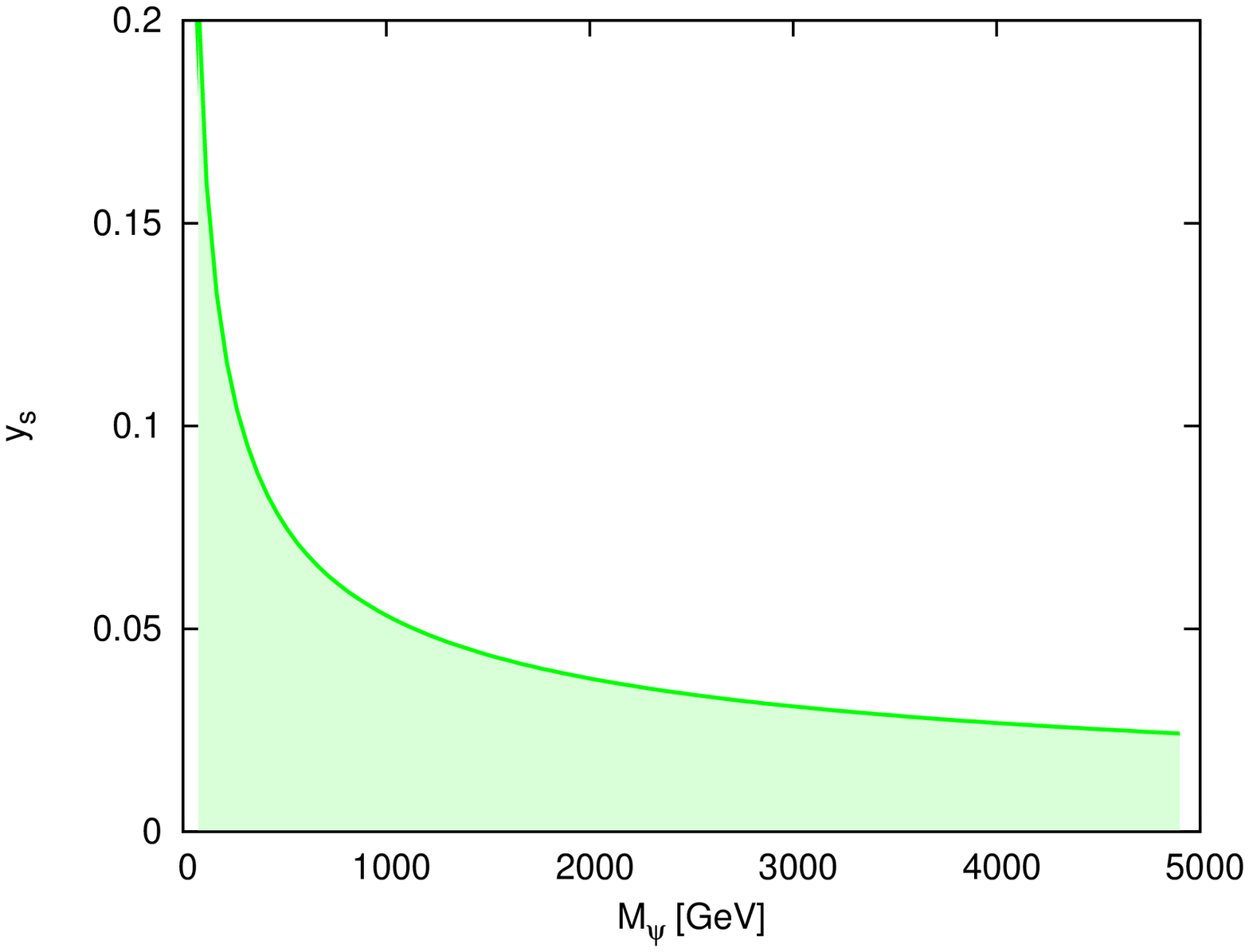,width=7cm}\hspace{0cm}\epsfig{figure=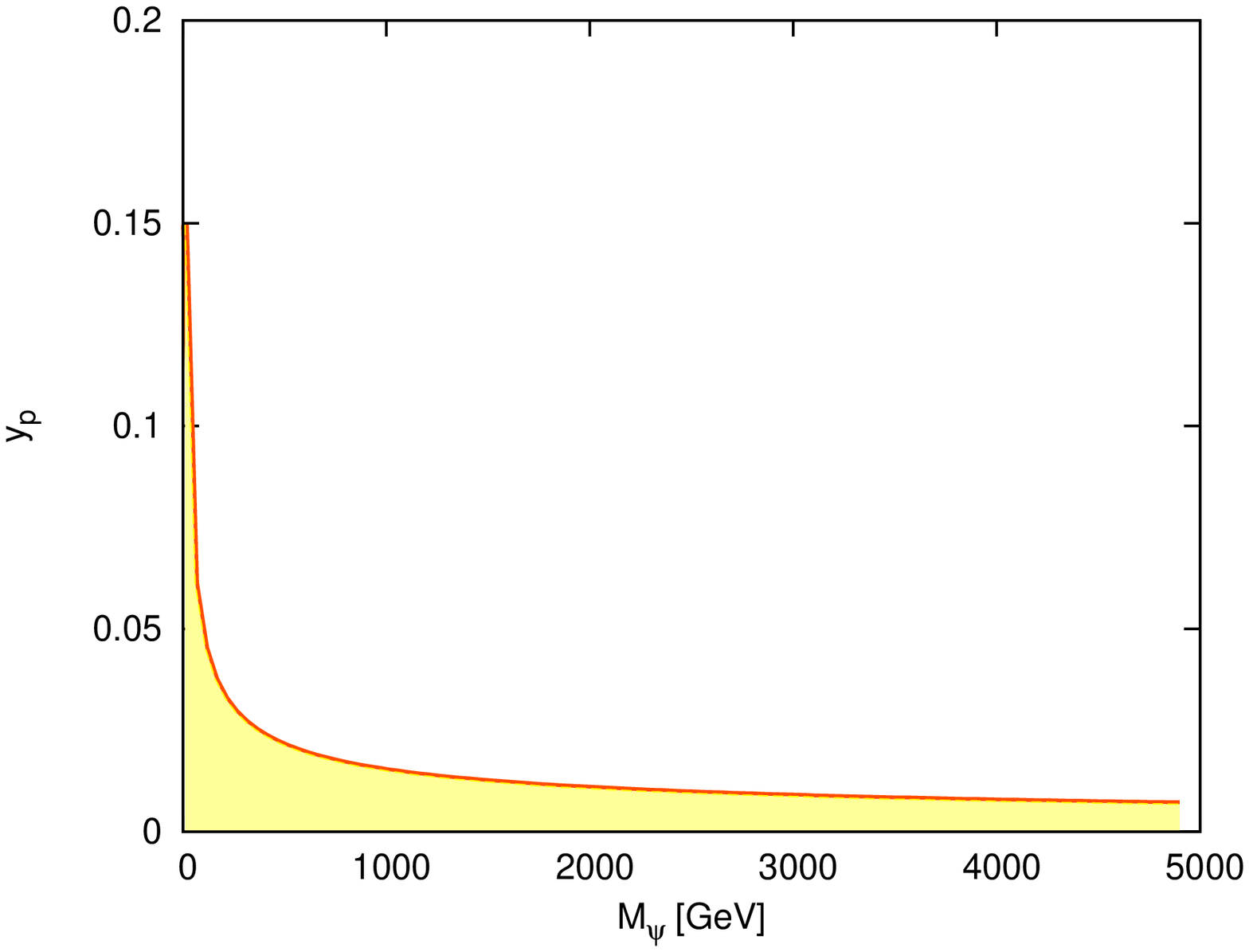,width=7cm}}
\centerline{\vspace{0.5cm}\hspace{0.5cm}(e)\hspace{6cm}(f)}
\centerline{\vspace{-0.7cm}}
\caption{Shadow areas depict allowed range in masses of mediators and couplings  for different vector DM models which are consistent with magnetic dipole moment of muon. The models~1-6, respectively correspond to Figs.~a-f.}\label{magnetic}
\end{center}
\end{figure}

In the following, we consider the discrepancy between experiment and the SM prediction for the magnetic moment of the muon which have been calculated in \cite{Bennett:2006fi}-\cite{Davier:2003pw}
\begin{equation}
\Delta a_\mu = (7.8\pm 10.48 ~{\rm to} ~22.1\pm 11.31)\times 10^{-10}, \label{deltamagnetic}
\end{equation}
where the error was combined of statistical, systematic and theoretical uncertainty. In this work, we consider above SM deviation and analyze the models contributions to the magnetic moment of the muon. The one-loop  contribution of VDMs to the magnetic moment of muon can be classified by\cite{Leveille:1977rc}-\cite{Queiroz:2014zfa}:
\begin{align}
{\text{model 1:}} \quad  &\Delta a_{\mu}^s = \left(\frac{m_{\mu}} {2\pi M_{\phi}}\right)^2\left\lbrace
      -\left[\frac{7}{12}+ \ln\frac{m_\mu}{M_{\phi}}\right] (\lambda_s)^2\right\rbrace, \\
{\text{model 2:}} \quad &\Delta a_{\mu}^{ps} = \left(\frac{m_{\mu}} {2\pi M_{\phi}}\right)^2
      \left\lbrace\left[\frac{11}{12} - \ln\frac{m_\mu}{M_{\phi}}\right] (\lambda_p)^2\right\rbrace,\\
{\text{model 3:}} \quad &\Delta a_{\mu}^V = \left(\frac{m_{\mu}} {2\pi M_{V}}\right)^2
      \left\lbrace\frac{1}{3} (g_s)^2\right\rbrace,\\
{\text{model 4:}} \quad &\Delta a_{\mu}^{aV} = \left(\frac{m_{\mu}} {2\pi M_{V}}\right)^2
      \left\lbrace-\frac{5}{3}(g_p)^2\right\rbrace,\\
{\text{model 5:}} \quad  &\Delta a_{\mu}^{\psi} =\left(\frac{m_{\mu}} {2\pi M_{X}}\right)^2\left\lbrace  \left[\frac{ M_{\psi} }{m_{\mu}} -\frac{2}{3}\right]y_{s}^2\right \rbrace,\label{magnetic-f1}\\
{\text{model 6:}} \quad  &\Delta a_{\mu}^{a\psi} =\left(\frac{m_{\mu}} {2\pi M_{X}}\right)^2\left\lbrace \left[ -\frac{ M_{\psi}}{m_{\mu}} -\frac{2}{3}\right] y_{p}^2\right\rbrace,\label{magnetic-f2}
\end{align}\label{formula magnetic}
where $m_{\mu}$ is the muon mass, $M_{\phi}$, $M_{V}$, $M_{\psi}$  are the  scalar, vector and spinor mediator and $M_{X}$ is DM mass.  $g_s$, $g_p$, $\lambda_s$, $\lambda_p$, $y_s$ and $y_p$ are couplings of the SM leptons with new fields in accordance with interaction terms of Eq.~\eqref{lagrangian}. Fig.~\ref{magnetic} depicts allowed range for each case in masses of mediators and couplings which are consistent with magnetic dipole moment of muon. Comparing Figs.~\ref{ann} and \ref{magnetic} shows even for loose hadronic uncertainty on anomalous magnetic moment of muon, models 5 and 6 are excluded. This is due to the fact that models~5 and 6 have fewer free parameters than models~1-4 and the conditions for satisfying the relic density bound is more complicated. In \cite{Feng:2019rgm}, it has been shown that without chiral violation, the contribution to magnetic moments of muon is not significant. Since in our model, we consider separately vector and axial interactions for the fermion exchange, the results seem to be different. In Fig.~3-e and 3-f, we depicted allowed range of parameters space for $\Delta a_\mu$ and they include part of parameters space in which $\Delta a_\mu$ is smaller than the bounds of Eq.~\ref{deltamagnetic}. In these figures, excluded region include parts of parameters space in which the contribution of new physics is negative ( see Eq~\ref{magnetic-f1} and \ref{magnetic-f2}). If we combine vector and axial interactions, the contribution of new interaction to magnetic moment of muon will be negative and it does not depend on mass of fermion.

\subsection{Collider Constraints} \label{sec3-2}
Constraints on leptophilic DM interaction come from several experiments at LEP,  LHC and neutrino beam facilities. Some of the strongest  bounds on leptophilic models thus stem from such searches:

1. Four-lepton processes $e^+ e^- \rightarrow l^+ l^-$ and di-lepton resonance searches
in $e^+ e^- \rightarrow l^+ l^- \gamma$ which are strongly constrained by LEP measurements. Searches in the framework of these process lead to following bounds on couplings of the models\cite{Freitas:2014pua}:
\begin{align}
{\text{model 1:}} \quad  & \lambda_s /M_{\phi} < 2.7 \times 10^{-4} {\rm~ GeV}^{−1}~~(M_{\phi} >200~{\rm~ GeV}), \nonumber\\
& \lambda_s /M_{\phi} < 7.3 \times 10^{-4}~ \rm~ GeV^{-1}~~(100~{\rm~ GeV} < M_{\phi} < 200~{\rm~ GeV}), \\ 
{\text{model 2:}}  \quad  & \lambda_p /M_{\phi} < 2.7 \times 10^{-4} {\rm~ GeV}^{−1}~~(M_{\phi} >200~{\rm~ GeV}), \nonumber\\
& \lambda_p /M_{\phi} < 7.3 \times 10^{-4}~ \rm~ GeV^{-1}~~(100~{\rm~ GeV} < M_{\phi} < 200~{\rm~ GeV}), \\
{\text{model 3:}} \quad  & g_s /M_V < 2.0 \times 10^{-4} {\rm~ GeV}^{−1}~~(M_V >200~{\rm~ GeV}) , \nonumber\\
& g_s /M_V < 6.9 \times 10^{-4}~ \rm~ GeV^{-1}~~(100~{\rm~ GeV} < M_V < 200~{\rm~ GeV}) , \\
{\text{model 4:}} \quad  & g_p /M_V < 2.4 \times 10^{-4} {\rm~ GeV}^{-1}~~(M_V >200~{\rm~ {\rm~ GeV}}) , \nonumber\\
& g_p /M_V < 6.9 \times 10^{-4}~ {\rm~ GeV}^{-1}~~(100~{\rm~ GeV} < M_V < 200~{\rm~ GeV}).
\end{align} 

\begin{figure}[!htb]
\begin{center}
\centerline{\hspace{0cm}\epsfig{figure=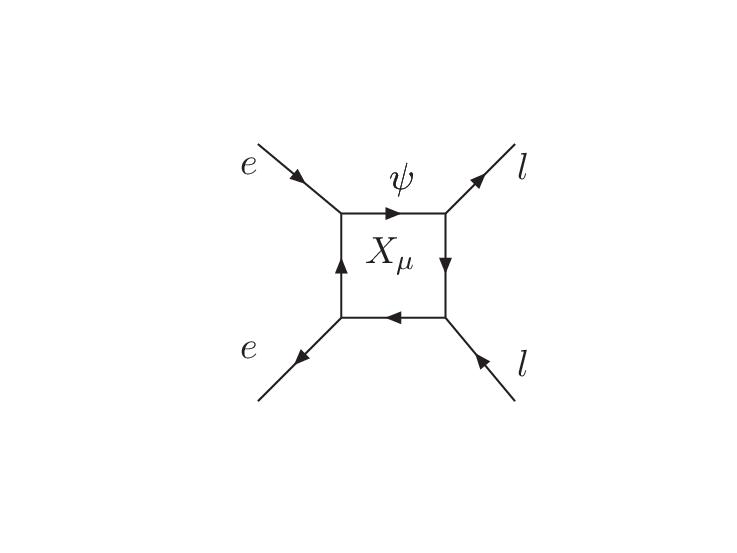,width=10cm}}
\centerline{\vspace{-2.5cm}}
\caption{Box diagram contribution to processes $e^+ e^− \rightarrow l^+ l^−$ for models 5 and 6}\label{box}
\end{center}
\end{figure}

The contributions of models 5 and 6 to process $e^+ e^- \rightarrow l^+ l^-$ arise from a box diagram which is depicted in Fig.~\ref{box}. Since this contribution for model 5 (6) is proportional to $y_s^4(y_p^4)$,  $M_{X}^{-4}$, and also DM mass ($M_{X}=1.4~\rm TeV$) is larger than the maximum LEP center-of-mass energy, LEP constraints on the model couplings will be irrelevant.
  
2. The production of a $\mu^-\mu^+$ from the scattering of a muon-neutrino with heavy nuclei (neutrino trident production: $\nu_{\mu}N \rightarrow \nu_{\mu} \mu \mu N$). 
Neutrino beam facilities, such as CHARM II collaboration \cite{Geiregat:1990gz} and the CCFR collaboration \cite{Mishra:1991bv} have been reported detection of trident events and quoted cross-sections in good
agreement with the SM predictions. These results strongly limit  vector mediator $V$ coupling to muons in model 3 if the $V$ couples to neutrinos $g_s\leqslant \frac{M_V}{1~\rm TeV}$\cite{trident}.

3. Searches for mono-photon events at $e^+ e^-$ colliders. This signature is characteristic for the process $e^+ e^- \rightarrow X \, X \, \gamma$. Since the LEP experiments did not observe an excess of mono-photon events beyond the expected background, a limit may be translated on leptophilic models\cite{mono-photonLEP}. 
It was shown in Ref.~\cite{mono-photonLEP}  that the bounds on four-lepton processes $e^+ e^- \rightarrow l^+ l^−$ exceed the limits from mono-photon searches at LEP  by about one order of magnitude. 

4. Drell-Yan production via an intermediate vector boson $V$ (in model 3 and 4) or scalar mediator produced (in model 1 and 2) as bremsstrahlung from a lepton at LHC. The mediator could subsequently decay
to combinations of lepton pairs and missing energy transverse (MET). In Ref.~\cite{LHC}, it was shown that the bounds on couplings are large when $M_V < M_Z$ and 
cross section falls rapidly with increasing mass of $V$. This means the constraints on couplings will be negligible for $M_{V}>100~\rm GeV$. 

\subsection{Direct Detection (DD) experiments} \label{sec3-3}
In this section, we will discuss the discovery potential of the models via direct DM searches. As it is mentioned, we consider the hypothesis that the vector DM particle $X$ couples directly only to leptons in particular the electrons but not to quarks.

\begin{figure}[!htb]
\begin{center}
\centerline{\hspace{0cm}\epsfig{figure=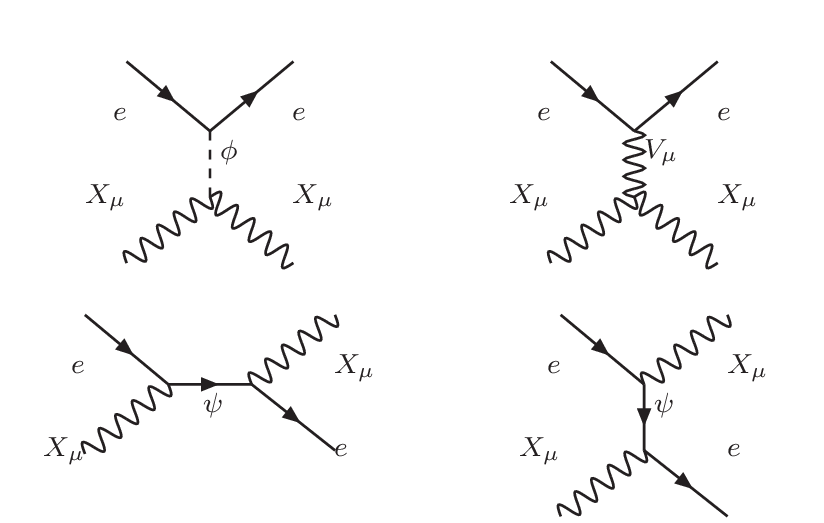,width=12cm}}
\caption{DM-electron vertexes for spinor, scalar and vector exchange.}\label{Direct Detection electron}
\end{center}
\end{figure}

Now we consider two types of interactions that arise when a “leptophilic” vector DM particle interacts in a detector:

1. Vector DM-electron scattering: In \cite{electrondirect} it was shown that a new class of of superconducting detectors which are sensitive to ${\cal{O}}(\rm MeV)$
electron recoils from DM-electron scattering. Such devices could detect DM as light as the warm DM limit, $M_{X} >∼ 1~ \rm keV$. In such experiment the whole recoil is absorbed by the electron that is then kicked out of the atom to which it was bound. In our model electron recoil can occur correspond to following Feynman diagrams (see Fig.~\ref{Direct Detection electron}).

2. Loop induced Vector DM-nucleus scattering: Although in our assumption DM couples only to leptons at tree level, an interaction with quarks is induced at 1 and 2-loop level, by coupling a photon to virtual leptons. This will lead to scattering of the
DM particle off nuclei.

In all cases that we assume, the interaction is induced by the exchange of an intermediate particle whose mass is much larger than the recoil momenta that is of
order a few MeV. Thus in  non-relativistic limit the elastic scattering cross section of the VDM with electron has following form\cite{DD}:
\begin{align}
{\text{model 1:}} \quad \sigma_{DM-e}& \approx\frac{g_{\phi}^2\lambda_s^2\mu_{eX}^2}{2\pi M_{\phi}^4} , \nonumber\\
{\text{model 2:}} \quad \sigma_{DM-e} & \approx\frac{g_{\phi}^2\lambda_p^2\mu_{eX}^2}{2\pi M_{\phi}^4} , \nonumber\\\
 {\text{model 5:}} \quad \sigma_{DM-e} & \approx\frac{y_{s}^4\mu_{eX}^2}{2\pi M_{X}^2M_{\psi}^2} , \nonumber\\
 {\text{model 6:}} \quad \sigma_{DM-e} & =0 , \label{Direct Detection electron1}
\end{align}
where $\mu_{eX}$ is the VDM-electron reduced mass. Since the models 3 and 4 can not explain DAMPE electron excess, we ignore them in this study. The last cross section is zero due to the odd number of $\gamma^ 5$ in the trace. We consider upper bound from the XENON100 experiment to search for DM interacting with electrons\cite{electron}. 
With no evidence for a signal above the low
background of such experiment, we can constrain parameters space of the models. For axial-vector interaction, it has been shown\cite{electron} that the cross-sections above $6 \times 10^{-35}~ cm^2$ for particle masses of $m_{X} = 2 ~\rm GeV$ is excluded. Eqs.~\eqref{Direct Detection electron1} predict that  
\begin{align}
\sigma_{DM-e}\approx\frac{g^4 m_{e}^2}{2\pi M_{mediator}^4} \approx g^4(\frac{M_{mediator}}{ 100~ {\rm GeV}})^{-4}\times 3\times 10^{-38} .
\end{align}
Therefore, even for a mediator mass of $1~ \rm GeV$ and general couplings $g=1$, the electron-DM cross section would be very smaller than the XENON100 \cite{electron} bounds. This feature depicts in Fig.~\ref{DDen}.  Note that DM-electron cross section is too small (e.g., $ \sigma_{DM-e} \lesssim 10^{-43} \, cm^2 $ for models 1 and 2) to constrain the models. However, these processes are more
important for DM masses below  ${\cal{O}}(\rm GeV)$, where the DM has insufficient kinetic energy to give detectable $(\rm keV)$ nuclear recoil energies.

\begin{figure}[!htb]
\begin{center}
\centerline{\hspace{0cm}\epsfig{figure=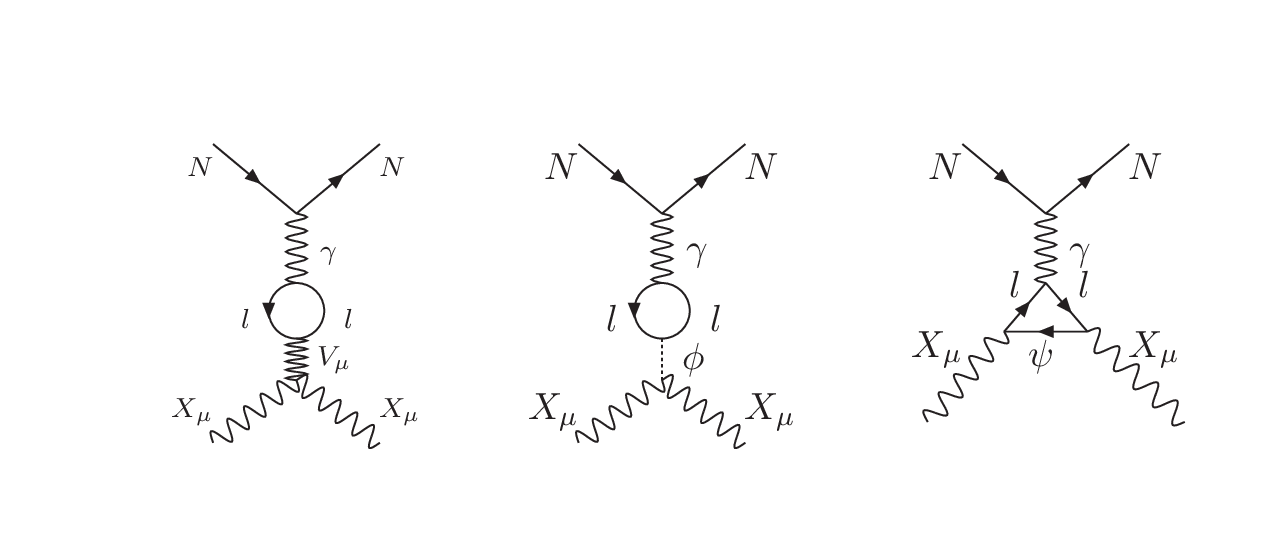,width=15cm}}
\centerline{\vspace{-1.2cm}}
\caption{DM-nucleus interactions by charged lepton induced and photon changed at 1-loop level for vector, scalar and spinor exchange. }\label{Direct Detection nucleus1}
\end{center}
\end{figure}  

\begin{figure}[!htb]
\begin{center}
\centerline{\hspace{0cm}\epsfig{figure=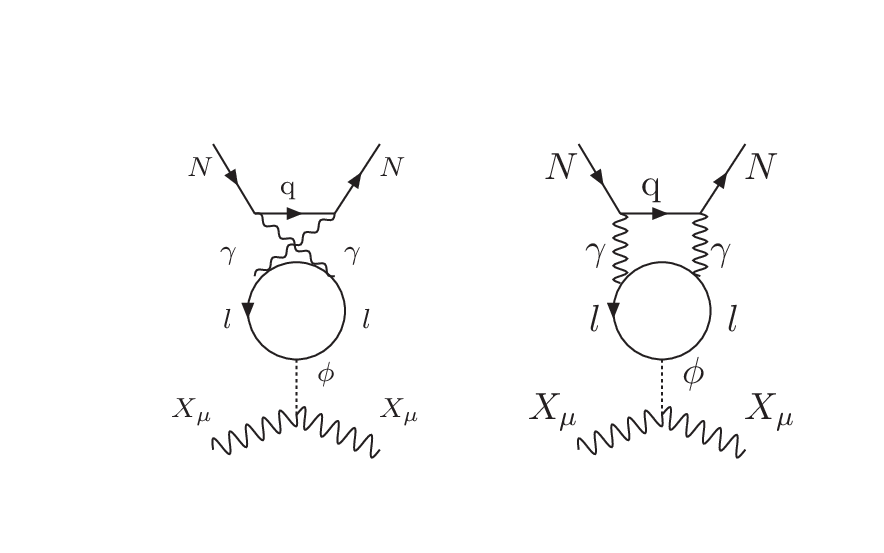,width=12cm}}
\centerline{\vspace{-1.2cm}}
\caption{DM-nucleus interactions by charged lepton induced and photon changed at 2-loop level for scalar exchange.}\label{Direct Detection nucleus2}
\end{center}
\end{figure}

As it was mentioned, leptophilic DM can scatter with quarks in DD experiments through lepton
loops. The leading DM-nucleus interactions
arise by charged lepton induced and photon changed at 1-loop level for spinor, scalar and vector exchange through diagrams of the form Fig.~\ref{Direct Detection nucleus1}. As it was discussed in \cite{DD}, for the models with scalar lepton current (model 1 and 2) in which low-velocity annihilation cross section requirement is satisfied, the one loop contribution involves the integral over loop momenta of the form: \begin{align}
 \int\frac{d^4q }{(4\pi)^4}{\rm Tr}[\Gamma\frac{q'\gamma^{\mu}+m_l}{q'^{2}-m^2_l}\gamma^{\nu}\frac{q\gamma^{\rho}+m_l}{q^{2}-m^2_l}] ,
\end{align}
where $\Gamma=1$ and $\gamma^5$ for model 1 and 2, respectively. The loop integral vanishes for these models, reflecting the fact that one cannot couple a scalar current to a vector current.
Since the models 3 and 4 (5 and 6) can not explain DAMPE electron excess (have been excluded by anomalous magnetic moment of muon), we ignore direct detection constraints for them.

\begin{figure}[!htb]
\begin{center}
\centerline{\hspace{0cm}\epsfig{figure=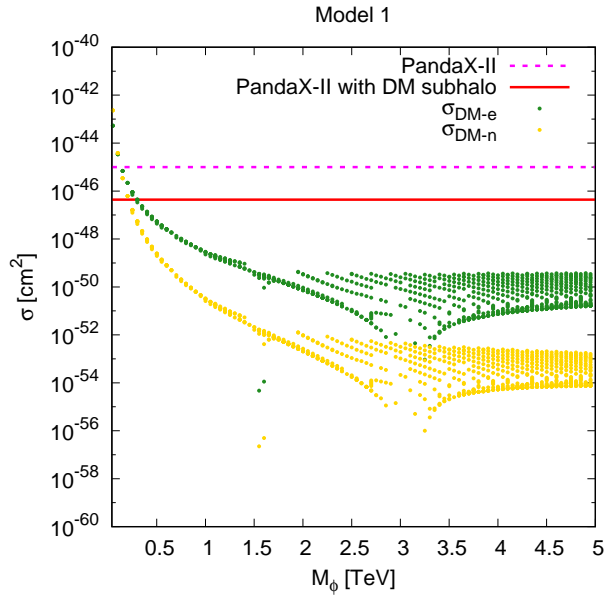,width=12cm}}
\centerline{\vspace{-1.2cm}}
\caption{DM-n and DM-e scattering cross sections versus scaler mediator mass.}\label{DDen}
\end{center}
\end{figure}

DM-nucleus interactions by charged lepton induced and photon changed at 2-loop level for scalar mediator are depicted in Fig.~\ref{Direct Detection nucleus2}. We calculate this contribution for model 1 which is given in following form:
\begin{align}
 \sigma_{DM-n}=\frac{\alpha_{em}^2Z^2\mu_{N}^2}{\pi^3A^2M^4_{\phi}}\sum_{\ell=e,\mu,\tau}(\frac{\pi\alpha_{em}Z\mu_{N}v}{6\sqrt{2}})^2(\frac{2g_{\phi}M_{X}\lambda_s}{m_lM_{\phi}})^2 , 
\end{align}
where $\alpha_{em}$ is the fine structure constant, $M_{\phi}$ is the mediator mass, $\mu_{N}\equiv m_{N} m_{X} /(m_N +m_{X})$
is the reduced mass of the DM-nucleus two particle system, $v = 0.001c$ is the velocity of the
DM near the Earth, $m_N$ , Z and A are the target nucleus mass, charge and mass number respectively. For model 2 (similar to 1-loop contribution), DM-n cross section is zero due to the odd number of $\gamma^ 5$ in the trace.

The best direct detection limits arise from the LUX \cite{LUX}, XENON1T \cite{XENON}, and
PandaX-II \cite{PandaX-II} collaborations. The PandaX-II  collaboration published the most stringent upper limit for a WIMP with mass larger than $100~\rm GeV$:
\begin{eqnarray} \label{constraints}
\rm{PandaX-II}: \sigma_{SI}\leq 8.6\times10^{-47}~cm^2\nonumber 
\end{eqnarray}
However, this limit is true for a DM local density of $ 0.3 \, \, \text{GeV}/cm^3 $. Since in DM explanation of DAMPE data, a nearby massive subhalo is assumed, therefore, it may influence the DM local density. For example, 
consider the NFW profile \cite{Navarro:1996gj} for the DM mass density in the subhalo,
\begin{equation}
\rho (r) = \rho_s \frac{(r/r_s)^{-\gamma}}{(1 + r/r_s)^{3-\gamma}} ,
\label{4-2}
\end{equation}
with $ \rho_s = 90 $ GeV/$ {\text{cm}^3} $, $ r_s = 0.1 $ kpc, and $ \gamma=1 $ .
If this subhalo is located at $ r = 0.17 $ kpc from the Earth, then, the induced local DM density in the solar system, will be $ 7.26 \, \, \text{GeV}/cm^3 $, which is $\sim 24$ times larger than local DM density inducd by Galaxy DM profile at solar system, i.e., $ 0.3 \, \, \text{GeV}/cm^3 $. This means we should consider direct detection limit 24 times more stringent. Regarding this new stronger bound, in Fig.~\ref{DDen} we have depicted allowed range in parameters space which are consistent with PandaX-II
direct detection experiment (for model 1).
As it has been seen, for $ M_{\phi} \lesssim 240 $ GeV the model 1 is excluded by PandaX-II direct detection experiment with DM subhalo contribution.
 
\subsection{Indirect detection} \label{sec3-4}
In addition to DAMPE, there are other constraints on the DM annihilation from other indirect detection experiments {\color{red}\cite{Belotsky:2019xti}} such as H.E.S.S. \cite{Abdallah:2016ygi}, FermiLAT \cite{Ackermann:2015zua} and IceCube \cite{Aartsen:2017ulx}.
For models 1 and 2, DM annihilation cross section which can explain DAMPE excess can simultaneously overcome all indirect detection constraint which are fairly weak for a DM mass of a TeV.

\section{Combined analysis} \label{sec4}
\begin{figure}
\begin{center}
\centerline{\hspace{0cm}\epsfig{figure=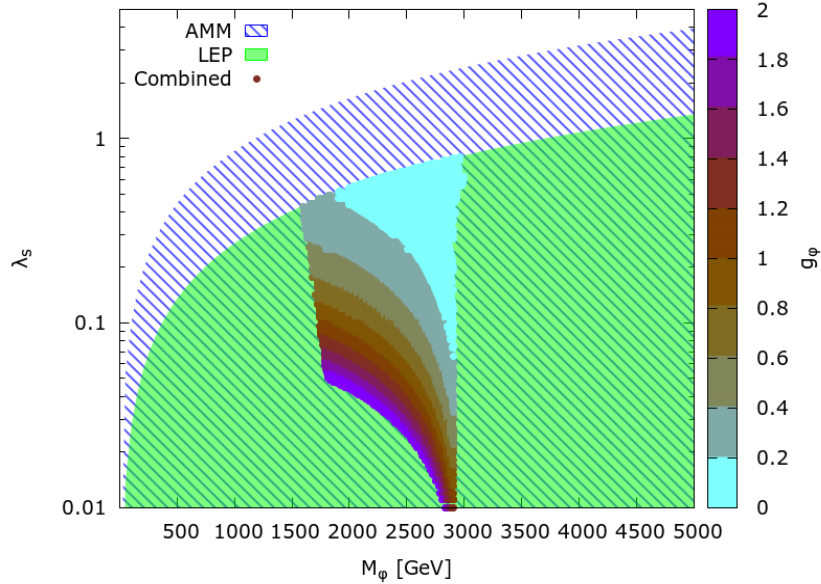,width=11cm}}
\centerline{\vspace{-0.7cm}}
\caption{The blue and green hatched areas depict regions for parameter space of the model~1 which are consistent with muon AMM and LEP constraints, respectively. The scatter points also satisfy relic density measurement, DAMPE excess condition ($ \left\langle \sigma v \right\rangle \simeq 
[2.2-3.8] \times 10^{-26} \, cm^{3}/s $), direct and indirect detection experiments.}\label{mod1}
\end{center}
\end{figure}

In this section, we present a combined analysis of all experimental constraints which were studied in previous section.   
 As it was mentioned before, model 3 and 4 are not considered because their DM annihilation cross section is too small and can not explain DAMPE excess. Model 5 and 6 are also excluded because the parameter space which can satisfy DM relic density can not simultaneously satisfy constraints from anomalous magnetic moment of muon. However, model 1 and 2 can survive all constraints. In Fig.~\ref{mod1}, we show regions which are consistent with relic density measurement,
DAMPE excess, direct and indirect detection experiments for different values of coupling in
model 1. For direct detection constraint, we consider PandaX-II direct detection experiment with DM subhalo contribution. Note that for anomolous magnetic moment (AMM), we consider stronger upper bound in Ref\cite{Davier:2003pw}. However (as it is seen), LEP constraint is stronger than AMM. Parameter space of the model satisfying $ \left\langle \sigma v \right\rangle \simeq 
[2.2-3.8] \times 10^{-26} \, cm^{3}/s $ is also depicted. It satisfies all other constraints including DM relic density, AMM, LEP, direct and indirect detection.

 According to Fig.~\ref{mod1}, LEP and DM annihilation cross section are the strong constraints that determine the parameter space of the model 1. Since, LEP constraint and DM annihilation cross section formulas for models 1 and 2 are similar, we only consider the model 1 in this section. Note that according to our study, there is not any direct detection experimental constraints on model 2. Therefore, the same result is true for this model.
 
\section{Conclusions} \label{sec5}

We have studied model independent leptophilic VDM candidates to determine which one of them can explain
the high energy electron-positron excess event recently observed by DAMPE experiment as well as other constraints from other DM searches.
The peak in the DAMPE electron-positron spectrum hints a nearby source for the
high energy electron-positron coming from the DM annihilation.
The peak is around 1.4 TeV, however, to account for the inevitable energy loss, we assumed a DM mass of about 1.5 TeV.
We have investigated all renormalizable interactions via a massive spin 0, 1/2, and 1 mediator between VDM and SM leptons.
We found that only two of six possible models can explain DAMPE excess, and, at the same time, survive all constraints including anomalous magnetic moment of muon, LEP, direct and indirect detection. In models 1 and 2, DM interacts with SM leptons via a scalar mediator. For $ M_{\phi} < 5000 $ GeV, we have scanned over parameter space, and found that if $ M_{\phi} \in [1500,3000] $ these models may explain DAMPE and simultaneously satisfy all experimental constraints.

\section*{Acknowledgments}
We would like to thank Dr. Karim Ghorbani for helping in micrOMEGAs code issues.

\end{document}